\documentclass[a4paper,11pt]{article}
\usepackage{jcappub}
\usepackage{lineno}
\usepackage{subcaption}
\usepackage{soul}

\newcommand{\nunubb}{2$\nu\beta\beta$}
\newcommand{\bb}{0$\nu\beta\beta$}
\newcommand{\Nnubb}{N$\nu\beta\beta$}

\newcommand{\Se}{$^{82}\mathrm{Se}$}

\newcommand{\ckky}{counts/keV/kg/yr}
\newcommand{\squaredsin}{$\sin^2{\theta}$}

\title{\boldmath Search for Sterile Neutrinos with CUPID-0}

\author[a]{O.~Azzolini}
\author[b]{J.W.~Beeman}
\author[c,d]{F.~Bellini}
\author[e]{M.~Beretta}
\author[e]{M.~Biassoni}
\author[e,f]{C.~Brofferio}
\author[g]{C.~Bucci} 
\author[e,f]{S.~Capelli}
\author[h,i]{V.~Caracciolo} 
\author[d]{L.~Cardani}
\author[e,f]{P.~Carniti}
\author[d]{N.~Casali}
\author[j]{E.~Celi}
\author[e,f]{D.~Chiesa}
\author[e]{M.~Clemenza}
\author[d,k]{I.~Colantoni}
\author[e]{O.~Cremonesi}
\author[d]{A.~Cruciani}
\author[g]{A.~D'Addabbo} 
\author[d]{I.~Dafinei}
\author[l,m]{S.~Di~Domizio}
\author[d,n]{F.~Ferroni}
\author[e,f]{L.~Gironi}
\author[o]{A.~Giuliani}
\author[g]{P.~Gorla} 
\author[e]{C.~Gotti}
\author[p]{L.~Gráf}
\author[a]{G.~Keppel}
\author[q,r]{J.~Kotila}
\author[s]{M.~Martinez}
\author[g,n]{S.~Nagorny}
\author[e,f]{M.~Nastasi}
\author[g]{S.~Nisi}
\author[t]{C.~Nones}
\author[g]{D.~Orlandi}
\author[g,n]{L.~Pagnanini}
\author[l,m]{M.~Pallavicini}
\author[e,f]{L.~Pattavina}
\author[e,f]{M.~Pavan}
\author[e]{G.~Pessina}
\author[c,d]{L.~Petrillo}
\author[d]{V.~Pettinacci}
\author[c,d]{S.~Pietrarota}
\author[g]{S.~Pirro}
\author[e]{S.~Pozzi}
\author[f,g]{E.~Previtali}
\author[g]{A.~Puiu}
\author[d]{A.~Ressa}
\author[g,u]{C.~Rusconi}
\author[v]{K.~Sch\"affner}
\author[d]{C.~Tomei}
\author[c,d]{M.~Vignati}
\author[t]{A.~S.~Zolotarova}

\collaboration{The CUPID-0 Collaboration}

\affiliation[a]{INFN Laboratori Nazionali di Legnaro, I-35020 Legnaro (Pd), Italy}
\affiliation[b]{Lawrence Berkeley National Laboratory, Berkeley, California 94720, USA}
\affiliation[c]{Dipartimento di Fisica, Sapienza Universit\`a di Roma, 00185 Roma, Italy}
\affiliation[d]{INFN, Sezione di Roma, 00185 Roma, Italy}
\affiliation[e]{INFN, Sezione di Milano - Bicocca, I-20126 Milano, Italy}
\affiliation[f]{Dipartimento di Fisica, Universit\`a di Milano - Bicocca, I-20126 Milano, Italy}
\affiliation[g]{INFN Laboratori Nazionali del Gran Sasso, I-67100 Assergi (Aq), Italy}
\affiliation[h]{Dipartimento di Fisica, Universit\`a di Roma Tor Vergata, I-00133 Roma, Italy}
\affiliation[i]{INFN, Sezione di Roma Tor Vergata, I-00133 Roma, Italy}
\affiliation[j]{Department of Physics and Astronomy, Northwestern University, IL, USA}
\affiliation[k]{Consiglio Nazionale delle Ricerche, Istituto di Nanotecnologia, c/o Dip. Fisica, Sapienza Università di Roma, 00185 Rome, Italy}
\affiliation[l]{INFN Sezione di Genova, I-16146 Genova, Italy}
\affiliation[m]{Dipartimento di Fisica, Universit\`a di Genova, I-16146 Genova, Italy}
\affiliation[n]{Gran Sasso Science Institute, 67100 L'Aquila, Italy}
\affiliation[o]{CNRS/CSNSM, Centre de Sciences Nucl\'eaires et de Sciences de la Mati\`ere, 91405 Orsay, France}
\affiliation[p]{Institute of Particle and Nuclear Physics, Faculty of Mathematics and Physics, Charles University in Prague, V Holešovičkách 2, 18000 Praha 8, Czech Republic}
\affiliation[q]{Finnish Institute for Educational Research, University of Jyväskylä, P.O. Box 35, 40014, Jyvaskyla, Finland}
\affiliation[r]{International Center for Advanced Training and Research in Physics (CIFRA), 409, Atomistilor Street, Bucharest-Magurele, 077125, Romania}
\affiliation[s]{Centro de Astropartículas y Física de Altas Energías, Universidad de Zaragoza, and ARAID, Fundación Agencia Aragonesa para la Investigación y el Desarrollo, Gobierno de Aragón,
Zaragoza 50018, Spain}
\affiliation[t]{IRFU, CEA, Universit\'e Paris-Saclay, F-91191 Gif-sur-Yvette, France}
\affiliation[u]{Department of Physics and Astronomy, University of South Carolina, Columbia, SC 29208, USA}
\affiliation[v]{Max-Planck-Institut f{\"u}r Physik, 85748 Garching, Germany}

\emailAdd{sylvie.pietrarota@roma1.infn.it}
\emailAdd{alberto.ressa@roma1.infn.it}

\abstract{Sterile neutrinos are well-motivated extensions of the Standard Model, introduced to address fundamental questions such as the origin of neutrino masses and the nature of dark matter. Exploiting the precise data reconstruction achieved by the CUPID-0 experiment, we searched for spectral distortions in the double $\beta$-decay of \Se\ compatible with the emission of a sterile neutrino. The analysis relies on the construction of a detailed background model down to 200 keV, enabling an accurate characterization of the main sources of contamination. Using a Zn\Se\ exposure of 9.95 kg$\cdot$yr, we explored sterile neutrino mass hypotheses between 0.5 MeV and 1.5 MeV. No evidence for a signal was observed in any scenario; therefore, we derived 90\% C.I. upper limits on the active-sterile mixing probability \squaredsin, obtaining the most stringent bound, \squaredsin\ $<8\times 10^{-3}$, for a sterile neutrino mass of 0.7 MeV.}

\keywords{massive sterile neutrino, double beta decay, scintillating bolometer, CUPID-0}

\begin{document}
\maketitle
\flushbottom

\section{Introduction}
\label{sec:intro}
Within the Standard Model (SM), neutrinos are described as massless, electrically neutral fermions that exist in three flavors and interact only through the weak force. The discovery of neutrino oscillations, however, revealed that they have finite masses and that flavor eigenstates are mixtures of mass eigenstates, providing clear evidence of physics beyond the SM. Although the minimal three-flavor oscillation framework successfully explains a wide range of experimental observations \cite{SNO,Homestake,Borexino,DayaBay,DoubleChooz,IceCube,Kamland,SuperK,Reno}, several anomalies in short-baseline oscillation experiments remain unresolved and suggest the existence of additional, undetected neutrino states \cite{LSND,MiniBooNE,gallium,ReactorAntinuAnomaly}. These hypothetical particles, known as sterile neutrinos, would not participate in SM interactions but could mix with the active flavors, thereby inducing small deviations from SM expectations in processes where active neutrinos are produced.

Sterile neutrinos are predicted in many extensions of the SM, such as the see-saw mechanism for mass generation \cite{Dasgupta_2021,Batra_2023}, and have also been proposed as viable dark matter candidates \cite{Abazajian_2017,Boyarsky_2019}. Motivated by these implications, their search has become an active and multidisciplinary endeavor, ranging from oscillation studies to astrophysical and cosmological probes \cite{inproceedings,Acero_2024}. Laboratory experiments play a crucial role, as they allow direct and model-independent investigations of their parameter space. In this context, nuclear processes such as double $\beta$-decay provide a unique opportunity. This is a rare second-order weak transition in which two neutrons in a nucleus simultaneously convert into two protons, emitting two electrons and two electron antineutrinos (\nunubb) \cite{PhysRev.48.512}. The hypothetical neutrinoless mode (\bb) is of particular importance: its observation would demonstrate the violation of total lepton number conservation and prove that neutrinos are Majorana particles \cite{PhysRev.56.1184,Primakoff_1959,10.1143/PTPS.83.1}. The existence of sterile neutrinos N coupling with the SM electron antineutrinos $\bar{\nu}_e$ can lead to two additional double $\beta$-decay final states, corresponding to the emission of one (\Nnubb) or two (NN$\beta\beta$) exotic particles \cite{Bolton_2021}:
\begin{align}
    \label{nuN} &(A,Z) \rightarrow(A, Z+2)+2e^-+\bar{\nu}_e+N \\ 
    \label{NN} &(A,Z) \rightarrow(A, Z+2)+2e^-+2N 
\end{align}
Provided that both decay channels are kinematically allowed, the total double $\beta$-decay rate would become an incoherent sum of three modes \cite{Bossio_2023, Bolton_2021}:
\begin{equation}\label{bb_gamma}
    \Gamma_{\beta\beta}=\cos^4{\theta}\,\Gamma_{\nu\nu}+2\cos^2{\theta}\sin^2{\theta}\,\Gamma_{\nu N}(m_N)+\sin^4{\theta}\,\Gamma_{NN}(m_N)
\end{equation}
where \squaredsin\ represents the mixing probability between electron and sterile neutrinos and $m_N$ is the sterile neutrino mass. Both quantities are free parameters of the theory, and the aim of this work is to constrain the ($m_N$, \squaredsin) parameter space by studying deviations from the standard \nunubb\ spectral shape. The contribution of the fully exotic mode, corresponding to the last term of equation \eqref{bb_gamma}, is strongly suppressed by a factor of $\sin^4{\theta}$, making it negligible for experimental searches.

\begin{figure}[htb!]
    \centering
    \includegraphics[width=0.67\linewidth]{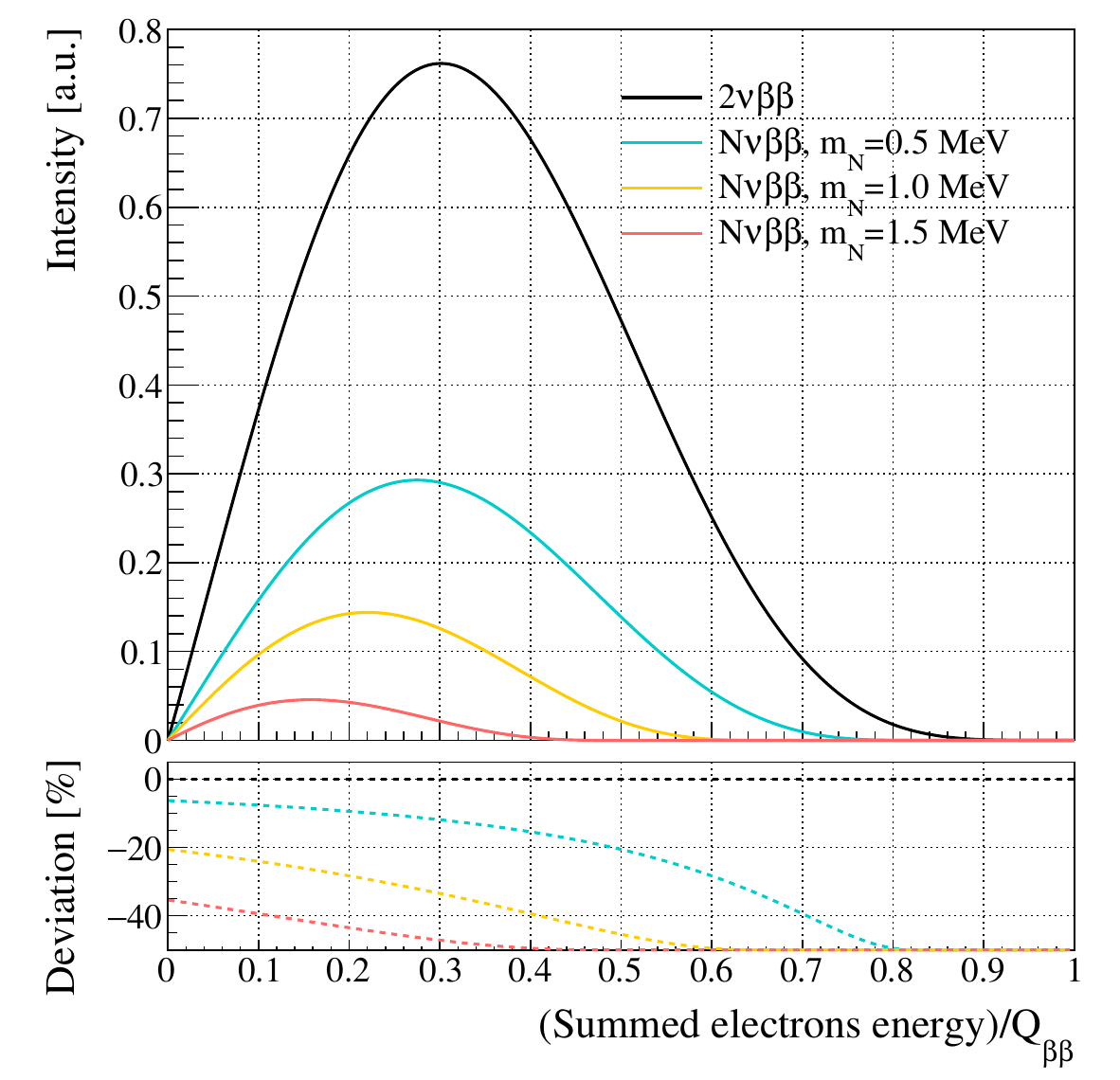}
    \caption{Comparison of the \nunubb\ spectrum with the predicted spectral shapes of \Nnubb\ for different sterile neutrino masses and assuming an active-sterile mixing angle \squaredsin=0.5. The distributions are normalized to the SM one and their analytical form is taken from Ref. \cite{Bolton_2021}. The bottom panel shows the deviation of the total double $\beta$-decay rate, including the sterile neutrino emission, from the purely SM decay rate. It can be observed that the distortion increases with the hypothesized sterile mass.}
    \label{fig:2vbb_vNbb}
\end{figure}

The summed kinetic energies of the electrons in the final state of \nunubb\ form a continuous spectrum ending at the Q-value ($Q_{\beta\beta}$) of the process. The phase-space (PSF) factor of the \Nnubb\ mode, being sensitive to the kinematics of the decay, would be affected by the presence of a massive fermion in the final state. This would lead to a characteristic distortion with respect to the \nunubb\ in the summed electron energy spectrum, which depends on the mass of the new state and on the active-sterile mixing angle. In particular, the endpoint of the distribution is shifted at $Q_{\beta\beta}-m_N$, resulting in a displacement of the peak of the distribution as well, as illustrated in Fig. \ref{fig:2vbb_vNbb}.

Experiments dedicated to the search for neutrinoless double $\beta$-decay are particularly well suited to investigate also other rare decay modes triggered by physics beyond the SM \cite{Blum_2018, 2019PhRvD..99i6005B, Deppisch_2020, Deppisch,devries2025scalarfuldoublebetadecay}. Indeed, to reach their primary goal they operate a large number of $\beta\beta$ candidate isotopes. Moreover, they must achieve excellent energy resolution, low background levels, and high detection efficiency, enabling detailed studies of spectral shapes. These requirements also make them powerful instruments for studying exotic decay modes such as those involving the emission of an additional neutrino species. 
The search for the \Nnubb\ decay mode has been performed on the $^{100}$Mo and $^{76}$Ge nuclei by the CUPID-Mo \cite{cupidMosterile} and GERDA \cite{Gerdasterile} experiments respectively. 
This paper presents a search for the emission of a sterile neutrino in the MeV mass range in the double $\beta$-decay of $^{82}$Se, performed with data from CUPID-0, the first large-scale demonstrator of the scintillating bolometer technology in the 0$\nu\beta\beta$ sector.

\section{The CUPID-0 experiment}\label{sec:cupid0}
The CUPID-0 experiment collected data from June 2017 to February 2020 at the Laboratori Nazionali del Gran Sasso (LNGS). Based on the experience gained in the field of scintillating cryogenic calorimeters for \bb\ searches \cite{pirro1,pirro2,pirro3,pirro4,Beeman_2013, Arnaboldi_2011}, CUPID-0 demonstrated the feasibility of this technique with a large-scale demonstrator.
This technology employs crystals containing a \bb\ candidate isotope that serve simultaneously as absorbers and signal sources. When the system operates at cryogenic temperatures, of the order of 10 mK, the temperature variation induced by a particle interaction becomes detectable ($\sim$100 $\mu$K per MeV of energy deposit). If the crystals are also good scintillators at low temperature, a fraction of the released energy is converted into light. Then, the simultaneous measurement of heat and scintillation light enables the discrimination between $\beta/\gamma$ induced signals and $\alpha$ particles --- being the latter the main background contribution in \bb\ search \cite{cupidbkgmodel,CUPID-Mo:bkgmodel,CUORE:2024fak}. 

CUPID-0 selected \Se\ as the candidate emitter, which features a Q-value ($2997.9\pm0.3$ keV \cite{Qvalue}) lying above the 2615 keV endpoint of $\gamma$-ray radioactivity, further mitigating background. The detector consisted of 26 ZnSe scintillating crystals arranged in five tower-like structures \cite{cupid0det}. Twenty-four crystals were enriched to (95.3±0.3)\% in $^{82}$Se \cite{82Se_enr} and used in this analysis, while the remaining two were made of natural ZnSe. The temperature increase in the absorbers was measured by Neutron Transmutation Doped Germanium (NTD-Ge) thermistors \cite{Hallerf}, glued to each crystal. The light detectors (LD) \cite{Beeman_2013} consisted of a Ge wafer, in which the photons were absorbed, coupled to a NTD-Ge, and were positioned between the ZnSe crystals along the tower. A VIKUITI{\texttrademark} plastic reflective foil surrounded each bolometer to enhance the light collection, ensuring a reflectivity greater than 98\% for wavelengths between 400-800 nm. Both crystals and light detectors were held in place by PTFE (polytetrafluoroethylene) clamps attached to the copper holder, which in turn connected the detector to the coldest thermal stage of a $^3$He/$^4$He dilution refrigerator working at a base temperature of 7.5 mK.

CUPID-0 successfully validated the effectiveness of the dual read-out bolometric technique by reaching a background level as low as $\sim$10$^{-3}$ \ckky\ at the \bb\ ROI, which allowed to set the best limit on $^{82}$Se \bb\ half-life \cite{cupidT120}. Moreover, thanks to the precise identification of the background sources affecting the energy spectrum up to 11 MeV and the excellent data reconstruction, the \nunubb\ of $^{82}$Se has been measured with unprecedented accuracy \cite{cupidT12}. 

\section{Background reconstruction} 
\label{sec:bkgmodel}
Understanding how the different contributions shape the observed energy spectrum is a fundamental task in rare-event search experiments such as CUPID-0. Building a detailed background model makes it possible to identify the origin of the main contaminants affecting the detector and to determine their activity. In particular, it is necessary to evaluate the contribution to the data of continuous spectral shapes, such as the one expected from the \Nnubb\ decay. 

In this work, we follow the methodology developed in Refs. \cite{Majoroni,Ressa:2025cji} and apply it to an updated set of Monte Carlo simulations. 
In addition, the background model is extended down to 200 keV, reaching lower energies than in previous CUPID-0 background studies.  
We analyze the data collected with the CUPID-0 detector from June 2017 to December 2018, corresponding to an active mass of 8.74 kg of Zn$^{82}$Se and a Zn$^{82}$Se exposure of 9.95 kg$\cdot$yr. 
The detailed data processing and selection are described in Ref. \cite{Azzolini_2018}. As a quantity of interest for this work, we only mention that the overall selection efficiency is constant at $\epsilon=(95.7 \pm 0.5)\%$ above 150 keV \cite{cupidbkgmodel}. 

The first step in constructing the background model is the analysis of the experimental spectra and the identification of the characteristic signature of  radioactive contaminants. These mainly originate from long-lived radioisotopes in $^{232}$Th, $^{238}$U and $^{235}$U decay chains, from $^{40}$K, and from cosmogenic activation, like $^{65}$Zn in ZnSe and $^{60}$Co and $^{54}$Mn in copper. Thanks to the passive shielding, the contribution due to environmental neutrons and $\gamma$-rays coming from the surrounding rocks can instead be considered negligible.

By exploiting the events topology within the detector and the $\alpha$ versus $\beta/\gamma$ discrimination, four distinct experimental spectra are defined to disentangle the background sources:
\begin{itemize}
    \item $\mathcal{M}_{1\beta/\gamma}$ is the energy spectrum formed by $\beta/\gamma$ events triggering a single crystal, covering the range from 200 keV to 5 MeV. The main component is the continuum produced by the 2$\nu\beta\beta$ decay of $^{82}$Se, on top of which the $\gamma$-ray lines of $^{65}$Zn, $^{40}$K and $^{208}$Tl decays are visible. 
    \item $\mathcal{M}_{1\alpha}$ consists of events identified as $\alpha$-particle interactions that triggered a single crystal, with energies between 2 MeV and 11 MeV.
    \item $\mathcal{M}_2$ and $\Sigma_2$ are the energy spectra of all events that triggered two bolometers within 20 ms. The choice of the time window was optimized by studying the time distribution of physical coincident events collected in the presence of a $^{232}$Th source \cite{PRL1cupid0}. $\mathcal{M}_2$ is built with the individual energies deposited in each crystal, while $\Sigma_2$ consists of the total energy released in the two bolometers. A minimum energy of 150 keV per crystal is required, and both spectra extend up to 11 MeV. These observables are especially sensitive to $\gamma$ backgrounds: high energy $\gamma$ rays can indeed cross more than one crystal, releasing a fraction of energy in their neighbors, so that some of the lines observed in $\mathcal{M}_{1\beta/\gamma}$ are also visible in $\Sigma_2$.
\end{itemize}
A total of 33 identified background sources are simulated with the Monte Carlo (MC) toolkit \textit{Arby}, a simulation package based on GEANT4 \cite{arby} (version 4.10.02). Following the results in Ref. \cite{ssd}, the \nunubb\ decay of $^{82}$Se is modeled according to the so-called Single-State Dominance approximation. Moreover, the recently determined corrections to the \nunubb\ spectrum \cite{niţescu2024radiativeexchangecorrectionstwoneutrino, morabit20242nubetabetaspectrumchiraleffective} are not considered here since their impact is expected to be subdominant with respect to the systematic uncertainties described in the following.

\begin{figure}
    \begin{subfigure}[c]{0.6\textwidth}
        \centering
        \includegraphics[width=\linewidth]{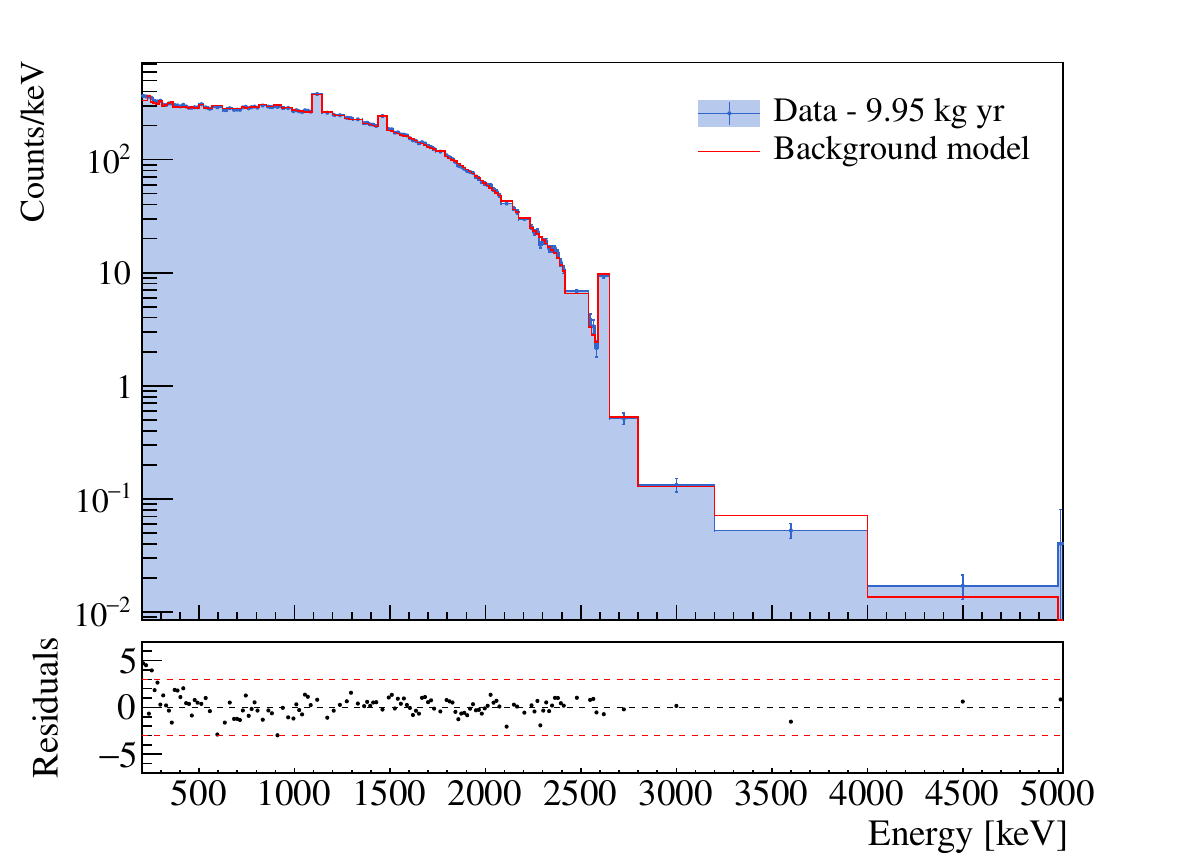}
    \end{subfigure}
    \begin{subfigure}[c]{0.39\textwidth}
        \centering
        \includegraphics[width=\linewidth]{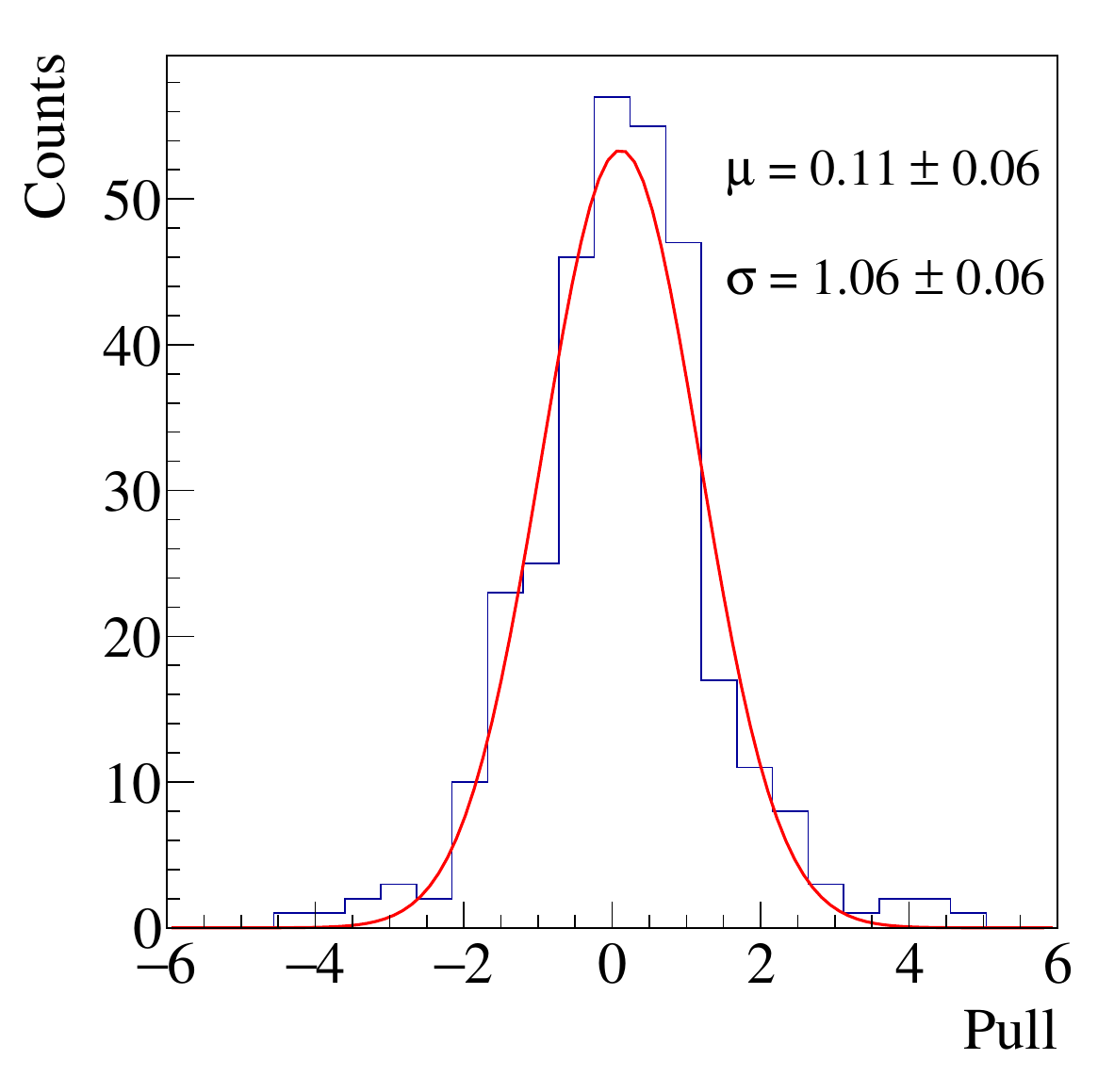}
    \end{subfigure}
    \caption{Validation of the background model in the $\mathcal{M}_{1\beta/\gamma}$ spectrum, using the full physics dataset of this analysis (Zn$^{82}$Se exposure of 9.95\,kg$\cdot$yr) and an energy threshold of 200 keV. \textit{Left}: Comparison between experimental spectrum and best-fit background model. The spectrum is shown with the variable binning adopted in the fit procedure. \textit{Right}: Total pull distribution, built from the normalized residuals of all bins in the four fitted spectra. The Gaussian fit gives $\mu=0.11\pm0.06$ and $\sigma=1.06\pm0.06$, with $\chi^2/\mathrm{d.o.f.}=21/17$.}
    \label{fig:bkgmodel}
\end{figure}

The same data selection cuts are applied to the MC templates. Each simulated background source is split into the four spectral classes $\mathcal{M}_{1\beta/\gamma}$, $\mathcal{M}_{1\alpha}$, $\mathcal{M}_{2}$, and $\Sigma_{2}$, so that they can be simultaneously combined to reproduce the corresponding experimental spectra.
The observed spectra are then compared with a linear combination of the simulated background components, whose coefficients are proportional to the unknown source activities. The coefficients, or scaling parameters, are extracted via a multivariate Bayesian fit (see also \cite{Majoroni}) whose methods are described in Appendix \ref{stats}.\\
To build the spectra, we adopted a variable binning strategy for two main reasons. Firstly, to mitigate mismatches between data and simulation in narrow structures (e.g.\ $\alpha$ peaks, $\gamma$ lines) arising from non-ideal detector responses \cite{Alduino_2017,jdhf-hn4l,cupidbkgmodel,Augier_2023}. Secondly, bins with less than 30 counts were merged 
to improve the stability of the data–MC comparison at higher energies. When neither of these conditions applied, a fixed bin width of 15 keV was used for the $\mathcal{M}_{1\beta/\gamma}$ spectrum, while 25 keV for the $\mathcal{M}_{2}$ and $\Sigma_2$ spectra because of their lower statistics. \\
The results of the fit to the experimental data for the $\mathcal{M}_{1\beta/\gamma}$ spectrum are reported in Fig. \ref{fig:bkgmodel}. The residuals are defined as the difference between the observed and the reconstructed number of events in each bin, normalized to the reconstruction uncertainty. The latter has been evaluated by combining the Poisson contribution from the data with the uncertainty on the reconstructed counts propagated from the model parameters covariance. The global pull distribution, given by the residuals across all bins in all spectral classes, is also shown in Fig. \ref{fig:bkgmodel}. Being the latter fully compatible with a standard Gaussian, it confirms that the model provides an adequate description of the data within statistical uncertainties.

\section{Sterile neutrino in double $\beta$-decay}
\label{sec:analysis}
If kinematically allowed, sterile neutrinos can be produced in any decay process involving Standard Model neutrinos through active-sterile mixing. In the \Nnubb\ decay of \Se, the kinematics requires the sterile neutrino mass to lie below the reaction Q-value, restricting the search to $m_N < Q_{\beta\beta} \simeq 3$ MeV. The actual $m_N$ values considered in this work are determined according to the sensitivity of the spectral
fit. In fact, the emission of a massive neutrino would shift both the endpoint and the
peak of the summed-electron energy distribution to lower energies. 
In particular, for $m_N$ higher than 1.5 MeV the peak of the distribution shifts below $\sim$500 keV, approaching the background model threshold. Conversely, for very small masses the spectrum becomes almost
indistinguishable from the SM case, again reducing the sensitivity. In this analysis, the investigated sterile neutrino masses span the range $0.5\, \mathrm{MeV} \leq m_N \leq 1.5\,\mathrm{MeV}$, sampled in 0.1 MeV steps. To evaluate their expected signatures, we simulated \Nnubb\ decays inside the ZnSe crystals of CUPID-0 following the formalism of Ref. \cite{Bolton_2021}.

For each value of $m_N$, the corresponding sterile-neutrino spectrum is added as an additional template in the background model and its own scaling parameter is fitted together with the ones of the other sources. This procedure yields the posterior probability distribution for the contribution of the sterile component to the observed data, proportional to the corresponding decay rate. The parameter of interest \squaredsin\ can then be written in terms of the \Nnubb\ and \nunubb\ decay rates and their respective phase-space factors:
\begin{equation}\label{eq:sin2}
\sin^2\theta
= \frac{G_{\nu\nu}}{2G_{\nu N}(m_N)}
\cdot \frac{\Gamma_{\nu N}}{\Gamma_{\nu\nu}}.
\end{equation}
In practice, the posterior distribution of \squaredsin\ is obtained by evaluating equation \eqref{eq:sin2} at each step of the MCMC sampling, thereby automatically accounting for correlations between the sterile and SM spectra. A flat prior in the interval [0,1] is assigned to the parameter \squaredsin.

The phase‑space factors for the \Nnubb\ decay mode are computed using the method of Ref. \cite{PSF}, with sterile‑neutrino effects included through the mass‑dependent modifications of Ref. \cite{Bolton_2021}. 
In this case, the available phase‑space is reduced by the sterile‑neutrino mass, effectively replacing $Q_{\beta\beta}$ with $Q_{\beta\beta}-m_N$. 
We note that, in the formalism of Ref. \cite{PSF}, the closure energy enters the PSF through the leptonic energy denominators, whereas in the sterile‑neutrino treatment of Ref. \cite{Bolton_2021} it appears only in the nuclear matrix element, with the PSF depending solely on kinematic quantities. The phase-space factors for the $^{82}$Se \Nnubb, computed here for the first time, are listed in Tab.~\ref{tab:PSF}. Their decrease with increasing $m_N$ implies progressively larger values of \squaredsin\ for a fixed decay rate. For comparison, the \nunubb\ PSF calculated within the formalism of Ref. \cite{Bolton_2021} is $G_{\nu\nu}=15.9\times10^{-19}\,\mathrm{yr}^{-1}$.
\begin{table}[htb!]
\centering
\begin{tabular}{c|c}
$m_{N}$ [MeV] & $G_{\nu N}(m_N)$ [$10^{-19}$yr$^{-1}$] \\
\hline
0.5 & 10.5\\
0.6 & 8.92\\
0.7 & 7.41\\
0.8 & 6.02\\
0.9 & 4.79\\
1.0 & 3.73\\
1.1 & 2.84\\
1.2 & 2.11\\
1.3 & 1.54\\
1.4 & 1.09\\
1.5 & 0.75\\
\end{tabular}
\caption{Phase-space factors for the $^{82}$Se \Nnubb\ decay as a function of the sterile neutrino mass, computed for the first time for this work.}
\label{tab:PSF}
\end{table}
Finally, we evaluated the systematic uncertainty of \squaredsin\ by varying the assumptions of the reference model and described in the previous section. Namely we repeated the fit:

\begin{itemize}
    \item using fixed bin widths (10 keV and 20 keV) along all the spectra;
    
    \item increasing the low-energy threshold at 300 keV. Higher thresholds were not considered because they would exclude significant parts of the analyzed spectra. This variation mainly affects the sensitivity to higher sterile masses, where the signal peak is shifted to lower energies; 
    
    \item shifting the energy scale of +3 keV and $-$5 keV, following the calibration bias studied by means of a $^{56}$Co source, reported in Ref. \cite{cupidres};
    
    \item alternatively removing degenerate sources of $^{232}$Th and $^{238}$U from the internal parts of the cryostat (labeled “No CryoInt" in Tab. \ref{tab:sterile_limits}), from the external ones (“No CryoExt") or from the lead shield (“No PbInt");
    
    \item removing sources that showed activities compatible with zero (“Minimal model");
    
    \item including the contribution of $^{90}$Sr-$^{90}$Y and $^{137}$Cs decay chains, whose contamination in the crystals are uncertain. Strontium-90 and Cesium-137 are anthropogenic contaminants produced by nuclear fission, whose activity is predicted to be similar \cite{Cs}. The first undergoes pure $\beta$-decay into Yttrium-90, which in turn also decays via $\beta$ emission, producing a broad and featureless spectrum extending up to 2.3 MeV. This shape is strongly degenerate with both the $2\nu\beta\beta$ and the \Nnubb\ spectra, increasing correlations among fit parameters. On the contrary, $^{137}$Cs decay chain, featuring a photo-peak at 661.7 keV, is better constrained by the model. Following the approach adopted by the CUORE collaboration \cite{CUORE_Sr}, we included both sources in the model constraining their activity to be the same.
\end{itemize}

\section{Results}
\label{sec:results}
For each sterile neutrino mass considered in this analysis, we performed a combined Bayesian fit across the four spectral classes $\mathcal{M}_{1\beta/\gamma}$, $\mathcal{M}_{1\alpha}$, $\mathcal{M}_2$, $\Sigma_2$, as outlined in appendix \ref{stats}, for the reference model and for each of the systematic tests described in Sec. \ref{sec:analysis}. In all cases, the data were found to be compatible with the background-only hypothesis. 
\begin{figure}
    \centering
    \begin{subfigure}[b]{0.49\textwidth}
        \centering
        \includegraphics[width=\linewidth]{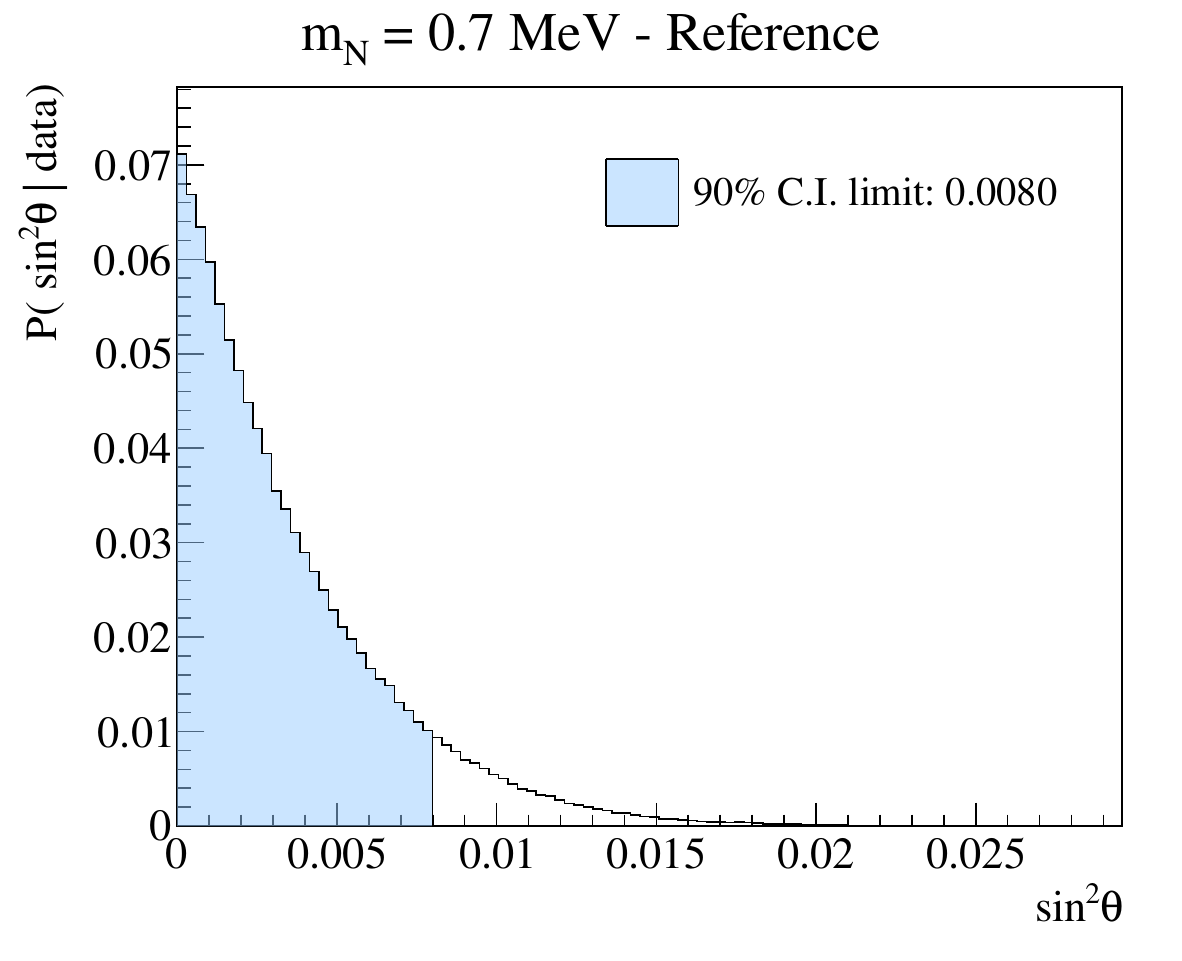}
        
    \end{subfigure}
    \begin{subfigure}
        [b]{0.49\textwidth}
        \centering
        \includegraphics[width=\linewidth]{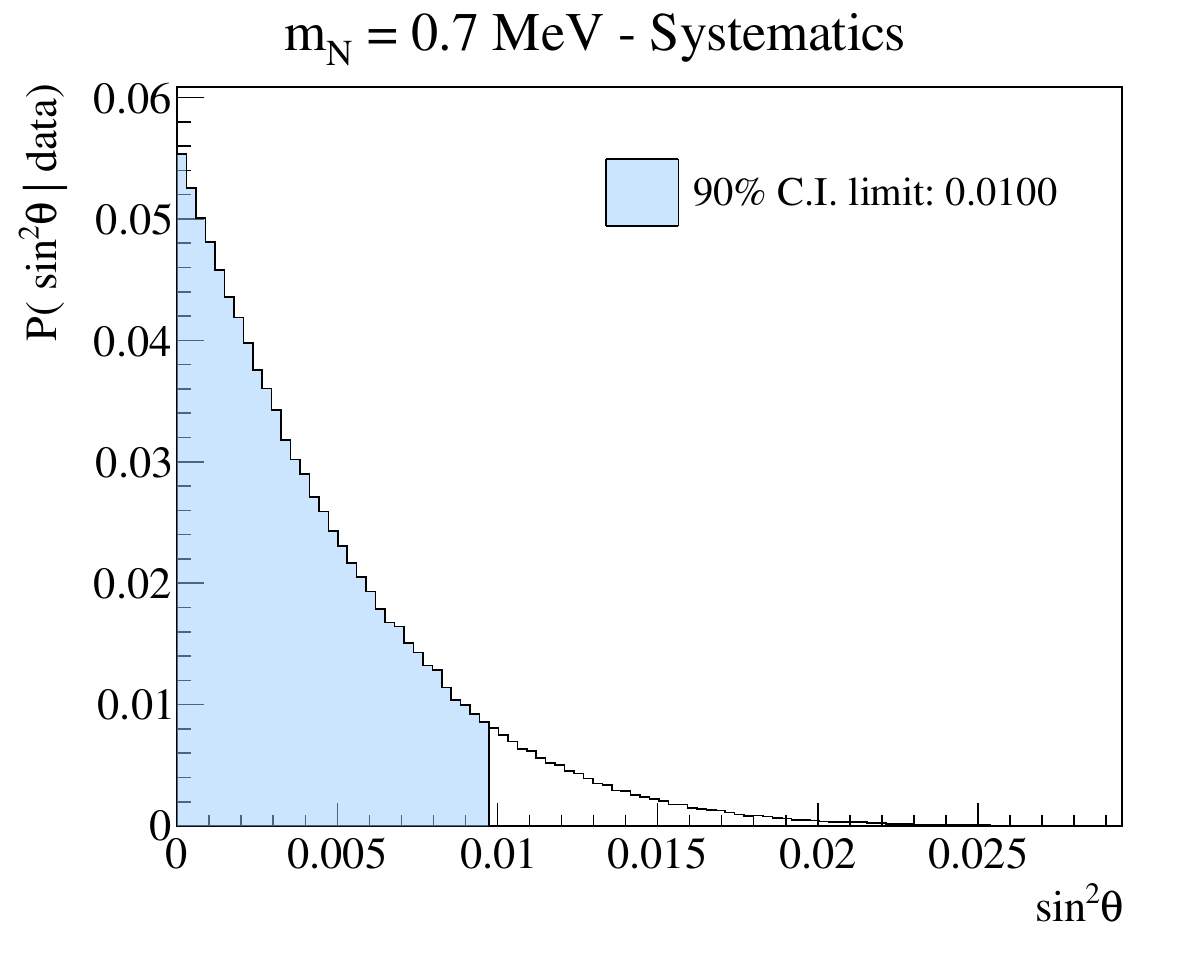}
        
    \end{subfigure}
    \caption{Posterior probability density for the mixing angle $\sin^2\theta$, obtained for a sterile neutrino mass hypothesis of $m_N = 0.7$ MeV, in the reference model (left) and including systematics effects (right). The shaded regions represent the 90\% credible intervals, from which the upper limits are derived.}
    \label{fig:sterile_posterior}
\end{figure}

To combine the results of the systematic tests, we applied the law of total probability, treating each test configuration as an alternative model for the data. In particular, we obtained the overall posterior by combining the individual model posteriors, weighted by their respective evidences, as described in appendix \ref{stats}, to take into account the data-model agreement over the different systematic tests. 

The resulting 90\% credible interval (C.I.) upper limits for each sterile neutrino mass, under both the reference and the systematic models, are summarized in Tab. \ref{tab:sterile_limits}. We observe that the mass range up to 1 MeV is mostly affected by the negative energy shift, while the energy threshold mainly impacts the higher sterile-neutrino mass values.

\begin{figure}
    \centering
    \includegraphics[width=\linewidth]{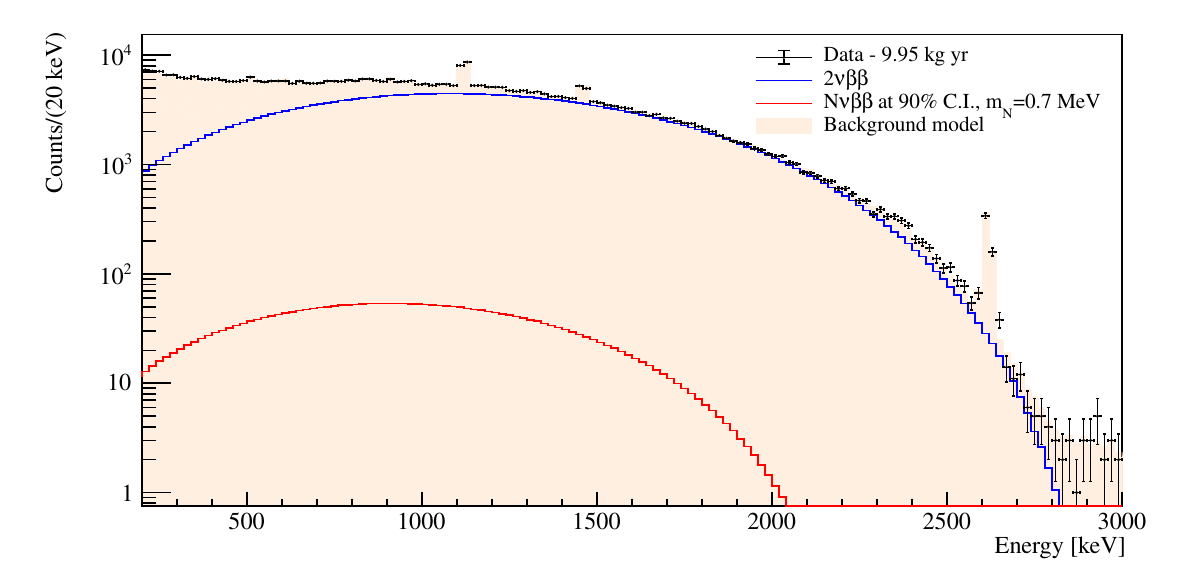}
    \caption{$\mathcal{M}_{1\beta/\gamma}$ spectra of experimental data, background model, 2$\nu\beta\beta$ and purely N$\nu\beta\beta$ decay with assumed $m_N=0.7$ MeV, obtained from the full physics dataset (Zn$^{82}$Se exposure of 9.95\,kg$\cdot$yr). The double $\beta$ simulated processes are scaled by the fitted mean and 90\% C.I. upper limit of the normalization coefficient, respectively. For illustrative purposes, all spectra are displayed using a fixed bin width.}
    \label{fig:bkg+N}
\end{figure}

In the reference case, the most stringent limit $\sin^2\theta<0.008$ was found for $m_N=0.7$ MeV, with comparable sensitivities up to 1.2 MeV. For higher masses the limits tend to weaken, because of the reduced sensitivity of the fit to their spectral shape, as well as decreasing phase-space factors as noted in Sec. \ref{sec:analysis}. As an example, Fig. \ref{fig:sterile_posterior} shows the posterior distributions of the parameter \squaredsin, with and without systematics, corresponding to a sterile neutrino mass $m_N=0.7$ MeV, with the 90\% C.I. highlighted. The contributions of the \Nnubb\ component for $m_N=0.7$ MeV, normalized to its 90\% C.I. upper limit, and of the $2\nu\beta\beta$ component, superimposed on the experimental $\mathcal{M}_{1\beta/\gamma}$ data, are illustrated in Fig.~\ref{fig:bkg+N}.

\begin{table}
    \centering
    \scriptsize 
    \setlength{\tabcolsep}{3.7pt}
    \begin{tabular}{l|lllllllllll}
         $m_N$ [MeV]&0.5&0.6&0.7&0.8&0.9&1.0&1.1&1.2&1.3&1.4&1.5 \\
         \hline
         \hline
         \textbf{Reference}& 0.0088&0.0083&0.0080&0.0084&0.0088&0.0100&0.0120&0.0159&0.0250&0.0497&0.1755\\
         \hline
         \hline
         bin=10 keV&+51\%&+53\%&+59\%&+42\%&+42\%&+32\%&+28\%&+19\%&+9\%&-8\%&-39\%\\
         bin=20 keV&+28\%&+34\%&+34\%&+31\%&+29\%&+31\%&+40\%&+41\%&+50\%&+66\%&+43\%\\
         \hline
         thr=300 keV&+17\%&+21\%&+24\%&+19\%&+22\%&+20\%&+28\%&+33\%&+46\%&+83\%&+102\%\\
         \hline
         +3 keV&-28\%&-26\%&-22\%&-20\%&-18\%&-15\%&-11\%&-6\%&-6\%&-2\%&-2\%\\
         -5 keV&+93\%&+89\%&+77\%&+60\%&+48\%&+39\%&+26\%&+20\%&+12\%&+6\%&-1\%\\
         \hline
         No CryoInt&-13\%&-10\%&-9\%&-10\%&-8\%&-9\%&-6\%&-4\%&-7\%&-9\%&-8\%\\
         No CryoExt&-14\%&-11\%&-8\%&-8\%&-4\%&-4\%&+1\%&+6\%&+9\%&+16\%&+33\%\\
         No PbInt&+8\%&+10\%&+8\%&+5\%&+4\%&+3\%&+4\%&+5\%&+2\%&+0.3\%&-5\%\\
         \hline
         Minimal model&+16\%&+15\%&+13\%&+12\%&+6\%&+1\%&+0.4\%&-0.2\%&-7\%&-15\%&-31\%\\
         \hline
         $^{90}$Sr-$^{90}$Y&+8\%&+10\%&+11\%&+9\%&+11\%&+9\%&+10\%&+10\%&+7\%&+4\%&-4\%\\
         \hline
    \end{tabular}
    \normalsize
    \caption{Limits on the $\sin^2\theta$ parameter at 90\% C.I. for the different neutrino mass and for the various models investigated, reference and systematics. The individual systematics results are expressed as percent variations with respect to the reference value. The labels in the first column correspond to the systematic configurations listed in Sec. \ref{sec:analysis}.}
    \label{tab:sterile_limits}
\end{table}

\section{Conclusions}
\label{sec:conclusions}
We performed the first search for sterile neutrino emission in the double-$\beta$ decay of \Se\ using the CUPID-0 data.
We built a background model, extended for this work down to 200 keV, through a Bayesian fit, enabling a robust extraction of the posterior probability distributions for the active–sterile mixing parameter $\sin^2\theta$ over the mass range 0.5--1.5 MeV. No evidence for a sterile-neutrino signal was found; therefore upper limits at the 90\% credible intervals were derived for each mass.
We evaluated systematic effects, including energy calibration, binning, thresholds, background composition, and spectral degeneracies.

\begin{figure}
    \centering
    \includegraphics[width=0.65\linewidth]{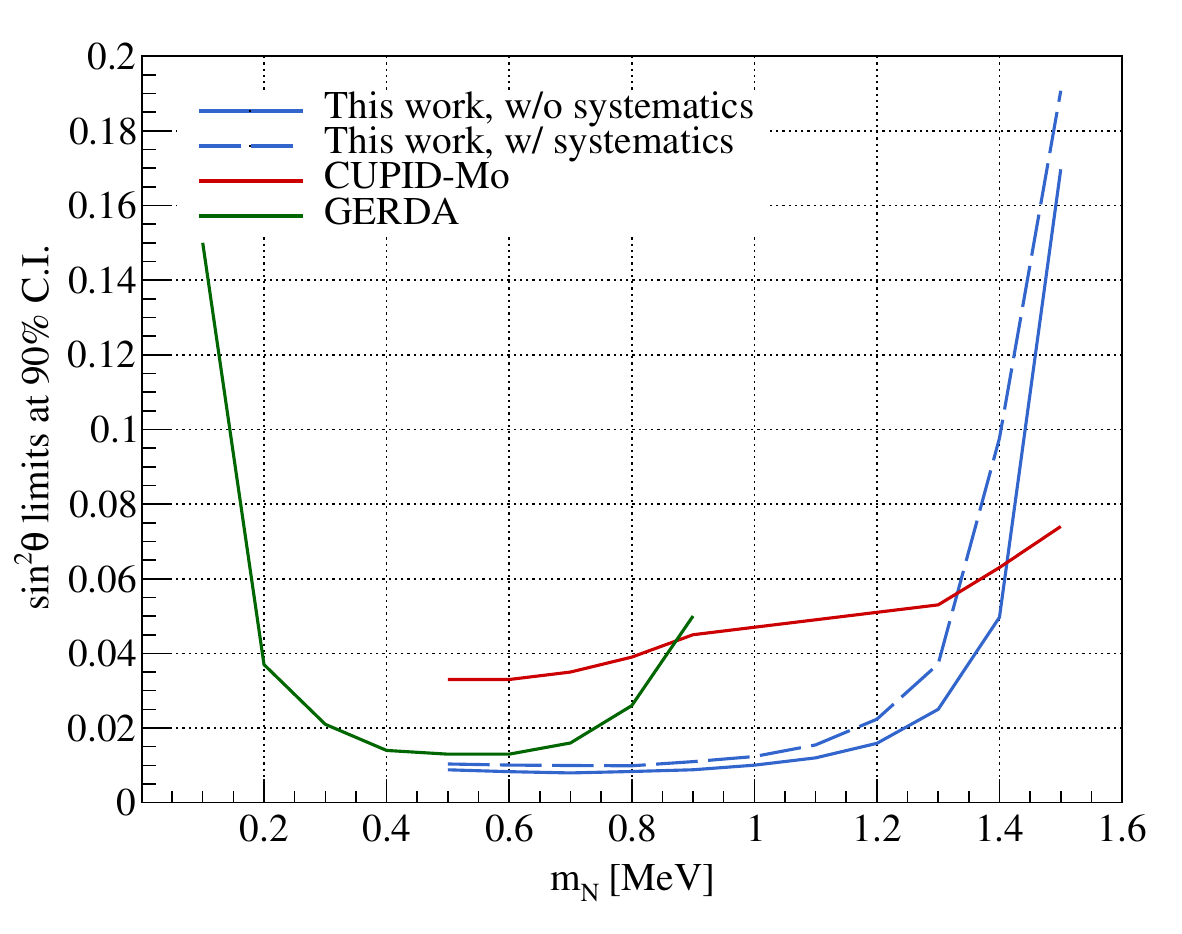}
    \caption{Limits at the 90\% C.I. on the mixing parameter $\sin^2\theta$ as a function of the sterile neutrino mass $m_N$. For comparison, we report the results obtained by other \bb\ experiments (CUPID-Mo and GERDA).}
    \label{fig:comparison}
\end{figure}

Similar searches for sterile neutrino emission in double $\beta$-decays of different isotopes have been performed by CUPID-Mo \cite{cupidMo} and GERDA \cite{gerdaexp}. CUPID-Mo, a demonstrator of the CUPID technology based on $^{100}$Mo, explored the same sterile neutrino mass range and, using a 1.5 kg$\cdot$yr exposure, set limits of $\sin^2\theta < 0.033$ at $m_N = 0.5$ MeV and $\sin^2\theta < 0.074$ at $m_N = 1.5$ MeV \cite{cupidMosterile}. GERDA performed a dedicated analysis in $^{76}$Ge, exploiting a 32.8 kg$\cdot$yr exposure, obtaining the strongest constraints around (500--600) keV with $\sin^2\theta < 0.013$ at 90\% C.L. \cite{Gerdasterile}. Compared to these results, the present analysis achieved stronger constraints across the explored mass range (see Fig. ~\ref{fig:comparison}). The improved sensitivity with respect to CUPID-Mo is mainly driven by the larger CUPID-0 exposure, while the advantage over GERDA originates from the higher $2\nu\beta\beta$ decay rate of $^{82}$Se, which enhances the statistical power of this search. 

Beyond double $\beta$-decay, tighter bounds on the active–sterile mixing in the MeV mass scale come from single $\beta$-decay studies and nuclear-structure tests. Spectral kink searches in the $^{20}$F $\beta$ spectrum constrain $\sin^2\theta < 1.8\times 10^{-3}$ for masses of (2--3) MeV \cite{beta_nulimits}, while precision measurements of superallowed $0^+ \to 0^+$ transitions exclude mixing down to $\sin^2\theta < (2.7\times10^{-4}$--$\,4\times10^{-4})$ in the range 1–9.4 MeV \cite{Ft}. Heavy sterile fermions in the MeV mass range could also decay into a lighter neutrino, an electron and a positron: the Borexino experiment performed a search for this signature in the neutrinos produced in the Sun via the $^{8}$B $\beta$-decay, obtaining limits $\sin^2\theta\leq(10^{-3}$--$4\times10^{-6})$ for $1.5\,\text{MeV}\leq m_N \leq 14$ MeV \cite{Bellini_2013}. \\
The results presented in this work provide a complementary limit to the mixing of hypothetical sterile neutrinos with the Standard Model neutrinos, testing their signature in $^{82}$Se \nunubb\ for the first time, proving the efficacy of CUPID-0 technology in the field of spectral shape studies.

\appendix
\section{Statistical framework}
\label{stats}
The reconstruction of the activities of the background sources, as well as of the signal strength, is based on a multivariate Bayesian fit. This method relies on constructing a likelihood function and defining a prior probability distribution for each scaling parameter $a_j$, in order to derive their posterior distributions. We consider the likelihood to be the product of Poisson probabilities over bins $i$ and spectral classes $\delta$:
\begin{equation}
    \mathcal{L}(\text{data}|\vec{a})=\prod_\delta\prod_{i=1}^{N_\text{bins}}\text{Pois}(N^\text{exp}_{i,\delta}|\,\lambda_{i,\delta}) \,\,\,\,\,\, \text{with}\,\,\, \delta\in\{\mathcal{M}_{1\beta/\gamma}, \mathcal{M}_{1\alpha}, \mathcal{M}_2, \Sigma_2\}
\end{equation}
where $N^\text{exp}_{i,\delta}$ is the number of observed counts in the $i$-th bin of the spectrum $\delta$. The dependence on the model parameters enters through the Poisson expectation value, i.e. the sum of the Monte Carlo simulation weighted by the unknown coefficients:
\begin{equation}
    \lambda_{i,\delta} = \sum_j a_j N^{j}_{i,\delta}\,\,.
\end{equation}
Here, $N^{j}_{i,\delta}$ denotes the number of simulated events from source $j$ in the $i$-th bin of spectrum $\delta$. 
The global posterior probability density of the scaling parameters is then given by the product of the likelihood and the priors $\pi(a_j)$
\begin{equation}\label{eq:posterior}
    P(\vec{a}|\text{data})\propto\mathcal{L}(\text{data}|\vec{a})\,\prod_j\pi(a_j)
\end{equation}
and is sampled via Markov Chain Monte Carlo (MCMC), generated using the \textit{Metropolis-Hastings} algorithm as implemented in the Bayesian Analysis Toolkit (BAT).\footnote{\href{http://mpp.mpg.de/bat/}{bat.mpp.mpg.de}} Uniform priors in physically allowed ranges were assigned to sources for which no independent information was available. Gaussian priors were instead adopted for contaminants whose activity had been previously constrained by independent measurements, as outlined in Ref. \cite{cupidbkgmodel}. 

From the global posterior in equation \eqref{eq:posterior} it is possible to derive the marginal posterior distributions of the individual components $a_j$, which can either be translated into background source activities or into the active-sterile mixing angle thanks to equation \eqref{eq:sin2}. In particular, the activity of each source can be computed as $A_j = a_j N_{j}/(\varepsilon M\Delta T)$, where $N_{j}$ is the total number of simulated decays, $\varepsilon$ the overall efficiency and $M\Delta T$ the exposure. Concerning the parameter of interest sin$^2\theta$, once its posterior distribution has been obtained for various models $M_s$, or systematic tests, these can be combined following the law of total probability:
\begin{equation}
    P(\text{sin}^2\theta | \text{data}) = \sum_s P(\text{sin}^2\theta | M_s, \text{data})P(M_s | \text{data})
\end{equation}
where $P(M_s | \text{data})$ is the posterior probability of the model $M_s$, computed as: 
\begin{equation}
    P(M_s | \text{data}) \propto P(\text{data} | M_s) \pi(M_s) = \int \mathcal{L}(\text{data}|\vec{a},M_s)\prod_j\pi(a_j) \text{d}a_j
\end{equation}
where we consider each model equally probable a priori ($\pi(M_s)$ = 1). The last expression is also known as $\mathit{evidence}$, and the integral runs over all the scaling parameters $a_j$ included in the specific model. 

\acknowledgments
This work was partially supported by the European Research Council (FP7/2007-2013) under Low--background Underground Cryogenic Installation For Elusive Rates (LUCIFER) Contract No. 247115. We are particularly grateful to M. Iannone for the help in all the stages of the detector construction, A. Pelosi for the construction of the assembly line, M. Guetti for the assistance in the cryogenic operations, R. Gaigher for the calibration system mechanics, M. Lindozzi for the development of cryostat monitoring system, M. Perego for his invaluable help, the mechanical workshop of LNGS (E. Tatananni, A. Rotilio, A. Corsi, and B. Romualdi) for the continuous help in the overall setup design. We acknowledge the Dark Side Collaboration for the use of the low-radon clean room. This work makes use of the DIANA data analysis and APOLLO data acquisition software which has been developed by the CUORICINO, CUORE, LUCIFER, and CUPID-0 Collaborations. This
work utilizes the ARBY software for GEANT4-based Monte Carlo simulations, which has been developed in the framework of the Milano-Bicocca R\&D activities and that is maintained by O. Cremonesi and S. Pozzi. L. Gráf acknowledges support from Charles University through project number PRIMUS/24/SCI/013. J. Kotila acknowledges the support from the EU-funded NEPTUN project (Project no. CF 264/29.11.2022 of the EU call PNRR-III-C9-2022-I8).

\bibliographystyle{JHEP}
\bibliography{biblio.bib}
\end{document}